\documentclass[twocolumn]{aastex63}
\usepackage[utf8]{inputenc}
\usepackage{soul}
\usepackage{xcolor}
\begin{document}
\title{Detection of OH 18-cm Emission from Comet C/2020 F3 NEOWISE using the Arecibo Telescope}
\author{Allison J. Smith}
\affiliation{Arecibo Observatory, Arecibo, PR, 00612, USA}
\affiliation{University of Central Florida, Orlando, FL, 32816, USA}
\author{D. Anish Roshi}
\affiliation{Arecibo Observatory, Arecibo, PR, 00612, USA}
\affiliation{University of Central Florida, Orlando, FL, 32816, USA}
\author{Periasamy Manoharan}
\affiliation{Arecibo Observatory, Arecibo, PR, 00612, USA}
\affiliation{University of Central Florida, Orlando, FL, 32816, USA}
\author{Sravani Vaddi}
\affiliation{Arecibo Observatory, Arecibo, PR, 00612, USA}
\affiliation{University of Central Florida, Orlando, FL, 32816, USA}
\author{Benetge B. P. Perera}
\affiliation{Arecibo Observatory, Arecibo, PR, 00612, USA}
\affiliation{University of Central Florida, Orlando, FL, 32816, USA}
\author{Anna McGilvray}
\affiliation{Arecibo Observatory, Arecibo, PR, 00612, USA}
\affiliation{University of Central Florida, Orlando, FL, 32816, USA}

\begin{abstract}

We report the detection of emission from the OH 18~cm $\Lambda$-doublet transitions toward Comet C/2020 F3 NEOWISE using the Arecibo Telescope. 
The antenna temperatures are 113$\pm$3~mK for the 1667~MHz line and 57$\pm$3~mK for the 1665~MHz line. The beam averaged OH column density (centered on the comet nucleus) derived from the 1667 transition is $N_{OH}$=1.11$\pm0.06\times10^{13}$~cm$^{-2}$.
We implemented the Haser model to derive an OH production rate. The estimated OH production rate using the 1667 transition is Q$_{OH}$=3.6$\pm0.6\times10^{28}$~s$^{-1}$, a factor of 2.4 lower than optically derived values for the same observing day, the difference of which is likely explained by quenching.
\end{abstract}


\section{Introduction} \label{sec:intro}
Discovered by astronomers associated with the $\it{NEOWISE}$ mission of the Wide-field Infrared Survey Explorer spacecraft \citep{mai11} on March 27, 2020, Comet C/2020 F3 (NEOWISE)\footnote{The Minor Planet Electronic Circular 2020-G05 is available at \url{https://minorplanetcenter.net/mpec/K20/K20G05.html}.} produced a vibrant show in the northern hemisphere during the summer of 2020, exciting both casual observers and astronomers alike. The comet reached perihelion on July 3, 2020 at a heliocentric distance of 0.3 AU, and continued on a trajectory that was favorable for Earth-based observations, presenting a unique opportunity to observe its composition and behavior \cite[see][for some early reports]{kni20,man20,lin20,kri20,dra20,smm+20}.

As comets spend most of their existence preserved in the icy outer Solar System, investigation of their astrochemistry is paramount for understanding the conditions when the Solar System was forming \cite[][and references therein]{ahe17}. Observations reveal a slew of molecules, likely released when the rapid temperature change incurred by the comet's approach toward the Sun causes water-ice to sublimate from the sub-surface nucleus into the coma \citep{des05}. That the molecular abundances vary for individual comets underscores the importance of gathering data on their chemistry when opportunities arise.

The hydroxyl (OH) molecule is the most common volatile and is generally used to assess the production of water as well as serving as a standard by which to compare other molecular abundances in cometary comae \citep{fel04}. \cite{bir74} first reported the detection of the 18 cm lines in comets. While electronic and vibrational transitions of OH are observable in the UV and infrared, the spectral resolution provided by radio observations of the hyperfine $\Lambda$-doublet transitions at 18-cm wavelength provides unique information on the velocity, distribution, and turbulence characteristics of the outflowing gas \citep{cro16}. The levels of the transition may be inverted or anti-inverted through UV pumping and fluorescence \citep{bir74, mie74, des81, sch88} and be manifest in observations as emission or absorption lines accordingly. Observing the $\Lambda$-doublet in cometary atmospheres often requires accounting for quenching of the inversion by collisions with electrons and ions, a phenomenon that has long been the subject of theoretical and observational investigations, and the neglect of which may result in underestimation of the OH production rate \cite[][and references therein]{cro02}.

The structure of this paper is as follows. We describe in \S~\ref{sec:obs} our observational setup and outline the steps taken to reduce the data in \S~\ref{sec:data}. We discuss the results in \S~\ref{sec:results} and the implications in \S~\ref{sec:dis}. Finally we conclude with a summary in \S~\ref{sec:Summary}.

\section{Observations} \label{sec:obs}
We observed Comet NEOWISE on July 31, 2020, using the 305-m radio telescope at the Arecibo Observatory (AO) in Puerto Rico, USA, during Director's Discretionary Time (Project A3470). We obtained the orbital parameters of the comet from the JPL HORIZONS system\footnote{\url{https://ssd.jpl.nasa.gov/horizons.cgi}} \citep{gio96} and present these in Table~\ref{tbl:obs}. We used the L-band Wide (LBW) receiver and configured the Wideband Arecibo Pulsar Processors (WAPPs) to observe simultaneously the OH main and satellite lines at 1665, 1667, 1612, and 1720~MHz. The total bandwidth for each subcorrelator board was 1.5625~MHz, each subdivided into 2048 channels, resulting in a spectral resolution of 0.7~kHz, i.e., 0.1~km~s$^{-1}$. As the LBW spectral baselines are stable on frequency scales comparable to the expected OH linewidths of the comet, we used the ON-only mode for our observations. The beam efficiency, $\eta_{MB}$, near 1667~MHz is $\sim$60\% and the half power beam width, $\theta_B \sim$ 2.9$\arcmin$. The mean zenith angle of the observations was 15$^\circ$. We observed a single pointing centered on the nucleus for the duration of the observations, approximately 110 min (consisting of 22~$\times$~5~min scans), and did not apply an on-line Doppler correction.

\begin{deluxetable}{lc}
\tabletypesize{\small}\tablecaption{Orbital Parameters of Comet NEOWISE.}
\label{tbl:obs}
\tablehead{ \colhead{Parameter\tablenotemark{a}} & \colhead{Value} \vspace{.2 cm}}
\startdata
UT range & July 31.781$-$July 31.875\\
Heliocentric distance ($r_h$) & 0.82 AU\\
Geocentric distance & 0.79 AU\\
Heliocentric velocity ($V_h$) & 37.17 km s$^{-1}$\\
Geocentric velocity & 34.92 km s$^{-1}$\\
Sun-target-observer (STO) angle & 78$^{o}$ \\
\enddata
\tablenotetext{a}{Tables values (aside from UT range) are quoted for the middle of the observing run.}
\end{deluxetable}

\section{Data Processing} \label{sec:data}

We processed these data using both existing\footnote{\url{http://www.naic.edu/~phil/software/software.html##idldoc}} and new routines developed with the Interactive Data Language (IDL) package. We first performed a quality check of the data by visual inspection of all 22 scans, each of which consists of 300 one-second integrations, and then averaged the two orthogonal linear polarization channels. There was no noticeable Radio Frequency Interference in our data. The antenna temperatures quoted are obtained using a calibrated noise source for which the average noise temperatures of the diodes were 8.48 and 9.35~K respectively, for the two polarizations. The shape of the bandpass was corrected for the calibrated spectrum by fitting a 3$^{\rm{rd}}$ or 4$^{\rm{th}}$ order polynomial to the line free channels and first subtracting and then dividing the spectrum by the polynomial. We then multiplied by the system temperature to retain the amplitude calibration. We obtained the final spectra by averaging the reduced data from all 22 scans and applying a velocity correction to account for the earth's rotation (217~m~s$^{-1}$). The source velocity changed by 245~m~s$^{-1}$ during the observations, and we used an FFT interpolation method to re-sample each calibrated spectrum during the averaging process in order to apply this correction. 

\section{Results} \label{sec:results}

\begin{figure*}
\begin{center}
\includegraphics[width=14cm,height=10cm]{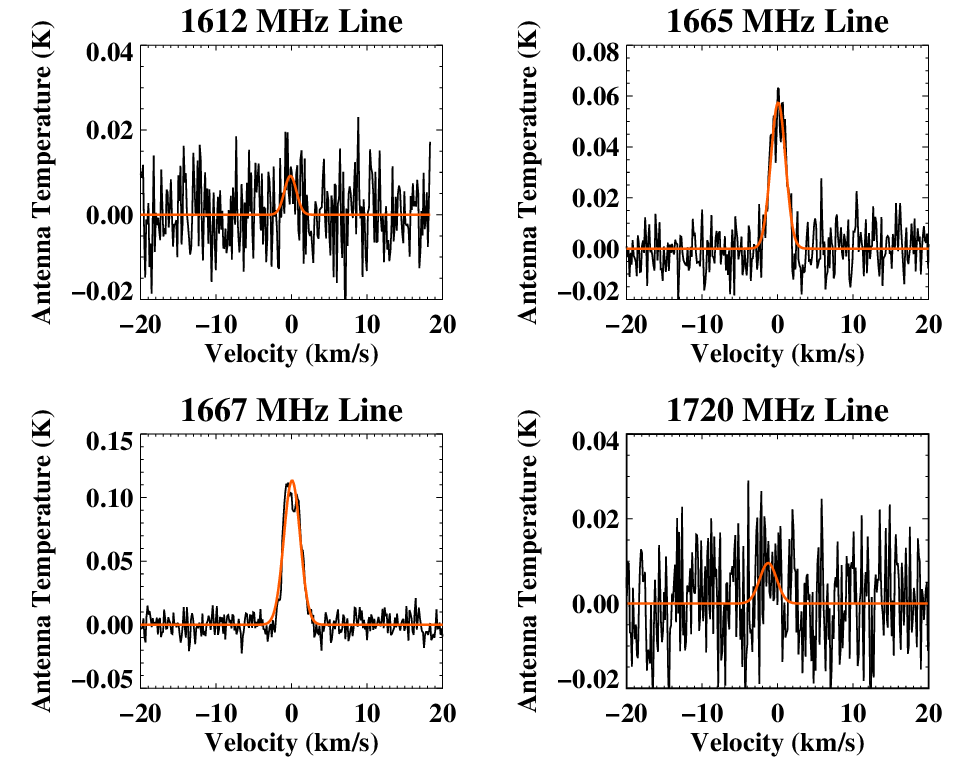}
\caption{Final averaged spectra for all four OH 18-cm lines in the nucleus-centered frame along with a fitted single-component Gaussian model (orange).} 
\label{fig1} 
\end{center}
\end{figure*}

These observations consist of a single observing session on Comet NEOWISE. While subsequent observations were planned, they were interrupted by the temporary shutdown of the Arecibo telescope due to the failure of one of its auxiliary cables supporting the telescope feed platform on August 10, 2020. Nevertheless, we detected relatively strong emission from the OH 18-cm main lines at 1665 and 1667~MHz during the first and only session. The line antenna temperature, $T_A$, spectral RMS, full-width at half maximum line width, $\Delta V$, and line central velocity with respect to the comet reference frame, $V_{CF}$, obtained with a single-Gaussian model (see Fig.~\ref{fig1}) are given in Table 2. The observed central velocities of the line are close to the expected line-of-sight velocity (34.92~km~s$^{-1}$) and are thus likely associated with the comet. We also used SIMBAD\footnote{\url{http://simbad.u-strasbg.fr/simbad/}} \citep{wen00} to confirm that there are no background OH sources (e.g., a molecular cloud) likely to produce a signal that could be confused with the signal from the comet.

 The detected line strengths in our observations are within the range of line amplitudes previously detected toward comets in Arecibo observations \citep{lov02} sampling similar physical scales at the source and heliocentric distances. The linewidths are also comparable, and the 1667:1665 line amplitude ratio is 1.98$\pm$0.12, marginally higher than the Local Thermodynamic Equilibrium (LTE) value of 1.8.  This implies the 1667~MHz line was likely crossing the transition from absorption to emission slightly ahead of the 1665~MHz line \citep[see][]{eli81}. The averaged spectra for all four lines are shown in Fig.~\ref{fig1}. Low-amplitude signals are visible at the expected velocity and LTE intensity for the satellite lines, though the peak amplitudes are only at the 1-$\sigma$ level.

\begin{deluxetable}{crccrc}
\tabletypesize{\small} \tablecaption{Parameters of the OH 18-cm lines}
\tablewidth{0pt} 
\label{tbl:widths}
\tablehead{ \colhead{Rest Freq.} & \colhead{$T_A$} & \colhead{RMS\tablenotemark{a}} & \colhead{$\Delta$ V} & \colhead{$V_{CF}$} & \colhead{$\int$ T$_L$ dV}\\
\colhead{(MHz)} & \colhead{(mK)} & \colhead{(mK)} & \colhead{(km/s)} & \colhead{(km/s)} & \colhead{(mK km/s)}}
\startdata
\hline
1612.2310 & 9(3) & 8.3 & 1.8(0.7) & $-$0.16(0.30) & 17(9)\\
1665.4018 & 57(3) & 7.6 & 2.3(0.1) & 0.08(0.05) & 139(10)\\
1667.3590 & 113(3) & 7.8 & 2.5(0.1) & 0.06(0.03) & 301(15)\\
1720.5300 & 10(4) & 11.6 & 2.7(1.1) & $-$1.24(0.48) & 29(18)\\
\hline
\enddata
\tablenotetext{a}{Corresponds to a velocity resolution of 0.14 km s$^{-1}$.}
\end{deluxetable}

\section{Discussion} \label{sec:dis}
\subsection{Haser Model and OH Production Rate \label{sec:hm}}

Observations of the $\Lambda$-doublet yield useful information on the total production rate of OH (Q$_{OH}$), the calculation of which requires knowledge of the OH distribution and excitation state. In order to estimate Q$_{OH}$, we implement the Haser Model, which assumes a spherical coma with the Haser density distribution for the daughter molecule (in our case, OH). We cross-check our results with the vectorial model \citep{com80} to assess their robustness and present alongside each other the parameters from both models following our description of the Haser calculation (see Table~\ref{tbl:model}).

The Haser density distribution depends on Q$_{OH}$, the daughter scale length ($l_d$), and the parent scale length ($l_p$), and requires a constant radial velocity for the daughter molecule ($v_d = |\vec{v}_d|$). Following \citet{des81}, we consider $l_p$ and $l_d$ scale with heliocentric distance as $r_h^2$. Q$_{OH}$ is a variable, which is determined by matching the model integrated line flux density with the observed value.

\begin{figure*}
    \centering
     \epsscale{1.15}\plottwo{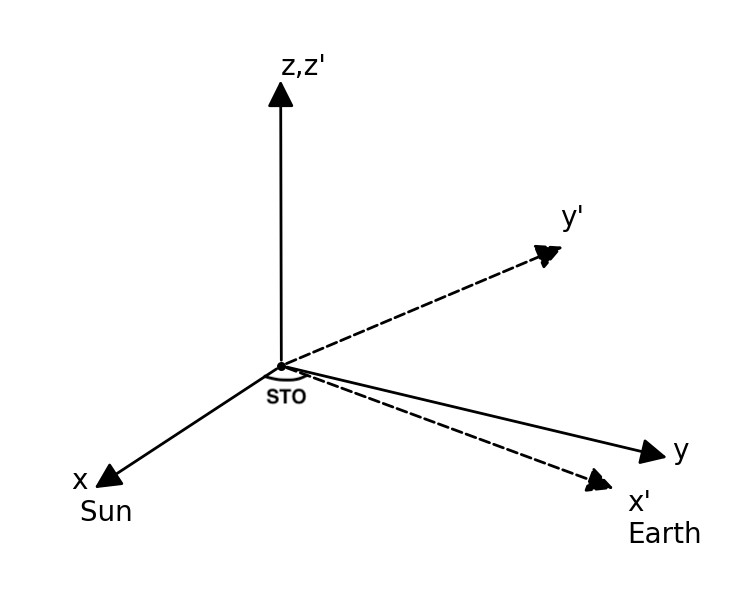}{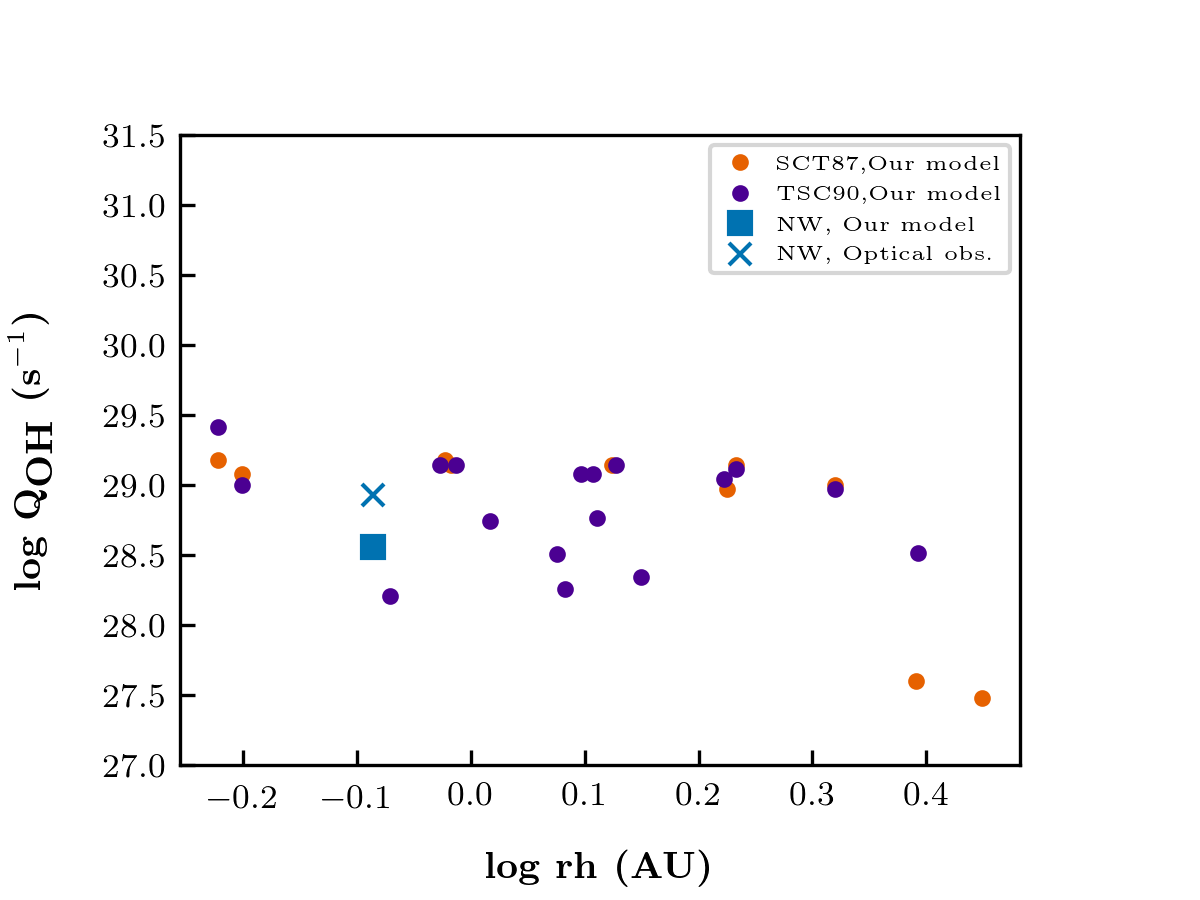}
    \caption{Left: The coordinate systems centered at the comet nucleus used in our model. The x$^{'}$ axis points toward Earth, while the x axis points toward the Sun. The two coordinates are rotated about the z-axis. The angle x-O-x$^{'}$ is the Sun-target-observer (STO) angle. {Right: Comparison of Q$_{OH}$ estimated for Comet NEOWISE with those obtained for other comets from literature. The Q$_{OH}$ values estimated using our model for the comet data from \cite{sch87} and \cite{tac90} are marked using filled circles. These values are within a factor of 2 of Q$_{OH}$ estimates given in \cite{sch87} and \cite{tac90}. The Q$_{OH}$ obtained for Comet NEOWISE is shown with filled square. The vertical width of the square is equal to the estimation error in Q$_{OH}$. For comparison, we also show the Q$_{OH}$ estimated for Comet NEOWISE from the optical observations (shown with blue cross).} The x-axis is the heliocentric distance in AU.}
    \label{fig3}
\end{figure*}

To compute the line amplitude, we consider a coordinate system (x$^{'}$-y$^{'}$-z$^{'}$ in Fig.~\ref{fig3}) placed at the center of the comet nucleus in which the x$^{'}$ coordinate points towards Earth. We use a second coordinate system (x-y-z in Fig.~\ref{fig3}) to compute the heliocentric velocities of the particles. The x-axis of this coordinate system points towards the Sun. The two coordinate systems are rotated about the z-axis and the angles x-O-x$^{'}$ is the Sun-target-observer (STO) angle. The space from 0 to $10 \times l_d$ is divided into N$_{grid}$ logarithmic intervals along x$^{'}$, y$^{'}$ \& z$^{'}$ coordinates. The density at the center of each cell in space is obtained from the Haser distribution, and the excitation state of OH is specified through the inversion parameter $i$ \citep[see][]{schleicher88}. It is well established that the excitation state oscillates between inverted ($i$ is positive) and anti-inverted ($i$ is negative) states with heliocentric velocity \citep{cro02}. Since the heliocentric velocity of the daughter molecule is the vector sum of the heliocentric velocity of the comet $\vec{V}_h$ and $\vec{v}_d$, we calculate it by summing up the x-component of $\vec{v}_d$ at the center of a given cell with $V_h = |\vec{V}_h|$. This velocity is then used to assign an inversion value to each cell.

Based on the basic assumption of the Haser model, the line emission from each cell when observed from Earth will fall in the interval $-v_d$ to $v_d$. Therefore, to obtain the line temperature as a function of velocity (i.e., to get the line profile), we divide the expected velocity interval (i.e., 2 $v_d$) into $N_v$ bins. Thus the velocity resolution in the simulation is $\Delta v = \frac{2 v_d}{N_v}$ and the corresponding frequency resolution is $\Delta \nu$. The observed velocity of emission from a cell would be the projection of $\vec{v}_d$ at the cell center in the direction of Earth. The line brightness temperature from a cell $T_L(v)$ in K for the optically thin case is \citep[see Eq 6 of][]{sch85}
\begin{equation}
    T_L (v) = C\; \left(T_{bg} + T_{L,bg}(v)\right)\; n_d\; i \; \Delta x^{'}\; \frac{1}{\Delta \nu},
\end{equation}
where $C=6.253 \times 10^{-10}$ cm$^2$ s$^{-1}$ for the 1667 transition, $T_{bg}$ is the background continuum temperature in K, $T_{L,bg}(v)$ is the contribution to the background temperature in K due to line emission from all the cells behind the cell under consideration at the velocity $v$, $n_d$ is the daughter density at the cell center in cm$^{-3}$, and $\Delta x^{'}$ is the line-of-sight thickness of the cell in cm. Assuming the line emission in a velocity bin is uniform, the line profile function is $\phi(\nu) = \frac{1}{\Delta \nu}$ and has units 1/Hz. The line brightness temperature vs. velocity in a grid parallel to the y$^{'}$-z$^{'}$ plane is obtained by summing the contributions from all the cells along the line of sight. The antenna temperature is then obtained by the weighted sum of the brightness temperature \citep[see Eq. 12 of][]{sch85}; the weights are determined by the FWHM observing beam, beam efficiency and the geocentric distance to the comet. The integrated line antenna temperature in units of K km s$^{-1}$ is the sum of the antenna temperature over the velocity bins multiplied by $\Delta v$. 

It is now established that when using the Haser model to estimate the production rate, equivalent values for the radial velocity and scale lengths need to be used \citep[see] []{com80}. The production rate estimated with `Haser equivalent' parameters compares well with predictions from, for example, the vectorial model \citep[see][]{boc90}. In our model, we used $l_{pH} = 2.4 \times 10^4$ km and $l_{dH} = 1.6 \times 10^5$ km \citep{ahe95}, where $l_{dH}$ and $l_{pH}$ are used to denote that they are equivalent values at 1AU, and scaled them by $r_h^2$ to get $l_p$ and $l_d$ for the computation. These values are the same scale lengths used for the estimation of the production rate from optical OH observations of Comet NEOWISE (Schleicher, D. private communication). We assumed that the OH ejection velocity is 1.0 km s$^{-1}$ \citep{boc90}. Using the trapezium line profile modeling method \citep{boc90}, we estimated the parent velocity as $1.12\pm0.05$ km s$^{-1}$. The Haser equivalent daughter velocity is then $v_{dH}= 1.49$ km s$^{-1}$ \citep[see] []{com80}, which is used in the model to get the density distribution. We found that only the \cite{schleicher88} inversion curve predicts an emission line at the heliocentric velocity of the observed comet. Therefore in our model we used their inversion values \citep[Table 5 of][]{schleicher88}. 

We used $N_v = 9$ and $N_{grid} = 200$ for the results presented in this paper. We found that the results are not affected by the grid size if $N_{grid} \ge 200$. We also compared the model predictions with computations made using a linear sampling of space and the results were found to be consistent with the logarithmic sampling.

The OH production rate obtained with our Haser model for Comet NEOWISE is $3.6\pm0.6 \times 10^{28}$ s$^{-1}$ (see Table~\ref{tbl:model}). In Fig.~\ref{fig3}, the production rate obtained for Comet NEOWISE is compared with rates obtained for other comets \citep[data taken from][]{sch87, tac90}. The integrated line intensity of the 1667 MHz transition was used for the estimation.  The values for heliocentric velocity, geocentric distance and STO used for this computation are listed in Table~\ref{tbl:obs}. The $T_{bg}$ towards the observed position is estimated as follows. The Galactic background temperature was obtained from the 408~MHz all-sky radio continuum survey \citep{has82} and the value after removing the Cosmic Microwave Background (CMB) contribution is scaled to 1667~MHz using a spectral index of $-2.6$. The estimated $T_{bg}$, including contribution from the CMB, is 3.23$\pm$.05~K. The $N_{OH}$, which represents the average over the Arecibo beam (centered on the comet nucleus), is estimated from the 1667 MHz transition to be $1.11\pm0.06 \times 10^{13}$ cm$^{-2}$.

We obtained an independent $Q_{OH}$ estimate of $4.4\pm0.07 \times 10^{28}$ s$^{-1}$from vectorial model calculations via Amy Lovell (Agnes Scott College, private communication). The vectorial $Q_{OH}$ and our $Q_{OH}$ are within the respective uncertainties of each model, and our $Q_{OH}$ is 82$\%$ of the vectorial value (see Table~\ref{tbl:model} for a comparison). 
The production rate estimated from optical OH line observations for the same observation date is 8.5 $\times 10^{28}$ s$^{-1}$ (Schleicher, D. private communication). We suggest that the difference between these two estimates could be due to the effects of quenching.
 
\begin{deluxetable}{llc}[htb!] 
\tabletypesize{\small} \tablecaption{Summary of Model Results}
 \tablewidth{0pt}
\label{tbl:model}
\tablehead{ \colhead{Input parameter} & \colhead{Q$_{OH}$} & \colhead{Note} \\
\colhead{values} & \colhead{$\times10^{28}$ s$^{-1}$} & \colhead{}} 
\startdata
\hline
$l_{dH}=1.6 \times 10^5$ km,   & $3.6\pm0.6$ & Haser equivalent \\
$l_{pH}=2.4 \times 10^4$ km, &  & model \\
$v_{p} = 1.12$ km s$^{-1}$ & & \\
$v_{d}=1.0$ km s$^{-1}$ & & \\
\hline
$\tau_{p}=8.2 \times 10^3$ s,   & $4.4\pm0.7$ & Vectorial model \\
$\tau_{d}=1.5 \times 10^5$ s, &   &  \\
$v_{d}=1.0$ km s$^{-1}$ &   &  \\
\hline
$l_{dH}=1.6 \times 10^5$ km,   & $8.5$ & Optical obs. \\
$l_{pH}=2.4 \times 10^4$ km   &   & \\
\hline
\multicolumn{3}{l}{Note $-$ $l_{dH}$, $l_{pH}$, the parent lifetime $\tau_p$, the daughter } \\
\multicolumn{3}{l}{lifetime $\tau_d$ are values at 1AU. The $v_p$ obtained from  } \\
\multicolumn{3}{l}{vectorial model is $1.22\pm0.07$ km s$^{-1}$ } \\
\enddata
\end{deluxetable}

\section{Summary} \label{sec:Summary}

We have observed emission lines from the OH 18-cm $\Lambda$-doublet transitions in Comet NEOWISE. We implemented a Haser model to derive the Q$_{OH}$. For the Haser equivalent model parameters, we got $Q_{OH} = 3.6\pm0.6 \times 10^{28}$ s$^{-1}$, which compares well with the values obtained with the vectorial model. The estimated OH column density $N_{OH} = 1.11\pm0.06 \times 10^{13}$ cm$^{-2}$. 
The derived $Q_{OH}$ is a factor of 2.4 lower compared to the optically derived value ($8.5 \times 10^{28}$ cm$^{-2}$), which is likely due to quenching.

\section{Acknowledgements} \label{sec:acknowledgements}
We are grateful for the Director's Discretionary Time dedicated to this project and for the assistance of Phil Perillat (AO) for his help with ephemeris tracking and telescope adjustments during the observations that led to high quality data. Amy Lovell gave extremely helpful input and provided an independent verification of the production rate using her vectorial modeling software. We very much appreciate valuable discussions with Dave Schleicher as well as the optical parameters he contributed. Chris Salter, Tapasi Ghosh, Yan Fernandez, Noem\'{i} Pinilla-Alonso, and an anonymous referee all gave very useful feedback on the manuscript. We made use of the SIMBAD database, operated at CDS, Strasbourg, France. The Arecibo Observatory is a facility of the NSF operated under cooperative agreement (\#AST-1744119) by the University of Central Florida (UCF) in alliance with Universidad Ana G. M\'{e}ndez (UAGM) and Yang Enterprises (YEI), Inc.

\pagebreak
\bibliography{neowise.bbl}
\bibliographystyle{aasjournal}
\end{document}